\documentclass[%
 aip,
 rsi,
 amsmath,amssymb,
 reprint,%
 superscriptaddress
]{revtex4-1}

\usepackage{graphicx}
\usepackage{dcolumn}
\usepackage{bm}

\usepackage[utf8]{inputenc}
\usepackage[T1]{fontenc}
\usepackage{mathptmx}
\usepackage{etoolbox}

\makeatletter
\def\@email#1#2{%
 \endgroup
 \patchcmd{\titleblock@produce}
  {\frontmatter@RRAPformat}
  {\frontmatter@RRAPformat{\produce@RRAP{*#1\href{mailto:#2}{#2}}}\frontmatter@RRAPformat}
  {}{}
}%
\makeatother

\begin{document}

\preprint{AIP/123-QED}

\title[A Milli-Kelvin Atomic Force Microscope Made of Glass]{A milli-Kelvin Atomic Force Microscope Made of Glass}

\affiliation{CAS Key Laboratory of Microscale Magnetic Resonance and School of Physical Sciences, University of Science and Technology of China, Hefei 230026, China}
\affiliation{Hefei National Laboratory, University of Science and Technology of China, Hefei 230088, China}
\affiliation{Anhui Province Key Laboratory of Scientific Instrument Development and Application, University of Science and Technology of China, Hefei 230026, China}

\author{Chengyuan Huang}
\affiliation{CAS Key Laboratory of Microscale Magnetic Resonance and School of Physical Sciences, University of Science and Technology of China, Hefei 230026, China}
\affiliation{Anhui Province Key Laboratory of Scientific Instrument Development and Application, University of Science and Technology of China, Hefei 230026, China}
\author{Zhenlan Chen}
\affiliation{CAS Key Laboratory of Microscale Magnetic Resonance and School of Physical Sciences, University of Science and Technology of China, Hefei 230026, China}
\affiliation{Hefei National Laboratory, University of Science and Technology of China, Hefei 230088, China}
\affiliation{Anhui Province Key Laboratory of Scientific Instrument Development and Application, University of Science and Technology of China, Hefei 230026, China}
\author{Mengke Ha}
\author{Haoyuan Wang}
\author{Qing Xiao}
\author{Changjian Ma}
\author{Danqing Liu}
\affiliation{CAS Key Laboratory of Microscale Magnetic Resonance and School of Physical Sciences, University of Science and Technology of China, Hefei 230026, China}
\affiliation{Anhui Province Key Laboratory of Scientific Instrument Development and Application, University of Science and Technology of China, Hefei 230026, China}
\author{Zhiyuan Qin}
\affiliation{CAS Key Laboratory of Microscale Magnetic Resonance and School of Physical Sciences, University of Science and Technology of China, Hefei 230026, China}
\affiliation{Hefei National Laboratory, University of Science and Technology of China, Hefei 230088, China}
\affiliation{Anhui Province Key Laboratory of Scientific Instrument Development and Application, University of Science and Technology of China, Hefei 230026, China}
\author{Dawei Qiu}
\author{Ziliang Guo}
\author{Dingbang Chen}
\author{Qianyi Zhao}
\author{Yanling Liu}
\author{Chengxuan Ye}
\author{Zhenhao Li}
\affiliation{CAS Key Laboratory of Microscale Magnetic Resonance and School of Physical Sciences, University of Science and Technology of China, Hefei 230026, China}
\affiliation{Anhui Province Key Laboratory of Scientific Instrument Development and Application, University of Science and Technology of China, Hefei 230026, China}
\author{Guanglei Cheng*}
\affiliation{CAS Key Laboratory of Microscale Magnetic Resonance and School of Physical Sciences, University of Science and Technology of China, Hefei 230026, China}
\affiliation{Hefei National Laboratory, University of Science and Technology of China, Hefei 230088, China}
\affiliation{Anhui Province Key Laboratory of Scientific Instrument Development and Application, University of Science and Technology of China, Hefei 230026, China}
\email{glcheng@ustc.edu.cn}

\date{\today}

\begin{abstract}
Milli-Kelvin atomic force microscopy (mK-AFM) presents an ongoing experimental challenge due to the intense vibrations in a cryogen-free dilution refrigerator and the low cooling power available at mK temperatures. A viable approach is to make the system exceptionally rigid and thermally insulating to decouple external vibrations and isolate heat dissipation from the piezo elements. Here, we present a low-cost and large scan-range mK-AFM that operates below 100 mK. All the essential parts of our mK-AFM, including the scanners, tip assembly, and microscope body, are custom-made of fused silica glass by taking advantage of its high specific modulus, extremely low thermal expansion coefficient, and excellent thermal insulation properties. We carefully balance the scan range (25 ${\mu}$m $\times$ 25 ${\mu}$m), heat dissipation, and stiffness of the system to reach optimal performance at mK temperatures. 
\end{abstract}

\maketitle


\section{\label{sec:level1}Introduction}

Since the invention of the scanning tunneling microscope (STM)\cite{binnig1982surface} and atomic force microscope (AFM)\cite{binnig1986atomic}, a broad branch of imaging tools collectively termed the scanning probe microscope (SPM) has proven to be very powerful across many research disciplines. Recent advancements in SPM are largely shaped by two dominant trends. The first involves developing novel imaging functionalities to probe diverse physical properties, including local magnetism\cite{baumann2015electron,marchiori2022nanoscale,aharon2022direct}, impedance\cite{barber2022microwave}, friction\cite{choi2011friction,thoren2016imaging}, and temperature\cite{halbertal2016nanoscale,halbertal2017imaging}, with enhanced sensitivity and resolution. For instance, by integrating the scanning superconducting quantum interference device (SQUID) on the tip of a sharp quartz pipette, the SQUID-on-tip microscopy employs the nano-SQUID sensor to deliver unparalleled sensitivity to temperature\cite{halbertal2016nanoscale,halbertal2017imaging} and spin\cite{aharon2022direct} with reasonably high spatial resolution. The second trend emphasizes operating at milli-Kelvin (mK) temperatures to address the intricate research demand on quantum material and topological phenomena at extremely low energy scales, including Majorana zero modes\cite{jack2021detecting} and edge state transport\cite{ji2024local}.

The ongoing challenges of achieving high-resolution mK imaging lie in several folds, primarily associated with the prevalent use of closed-cycle dilution refrigeration in the scientific community to avoid liquid helium consumption. The cooling power of a typical dilution refrigerator at 100 mK ranges from a few hundred microwatts to a few milliwatts\cite{zhao2019cryogenic}. However, the effective cooling power delivered to the sample can easily be an order smaller depending on the construction of the microscope. This requires careful heat management by restricting the usage of high-force piezoelectric elements, typically with large capacitance. Moreover, closed-cycle cryocoolers, such as pulse tubes and Gifford-McMahon cryocoolers, generate intense vibrations in the frequency range below 1 kHz, which severely interferes with SPM imaging.

Despite decades of efforts, very few SPM implementations have successfully demonstrated high-resolution imaging at mK temperatures while maintaining a relatively large scan area needed for generic AFM imaging. A common strategy involves mechanically decoupling the pulse tube cryocoolers from the cooling plates to mitigate vibrations\cite{pelliccione2013design}. In addition, multi-stage spring isolations can be employed to further reduce vibrations to tens of picometers at the expense of cooling power\cite{den2014atomic}. Notably, several homemade mK-STMs have been developed in dry dilution fridges and achieved atomic resolutions. Since the typical scan range of STM is around 1 ${\mu}$m $\times$ 1 ${\mu}$m, the scanner, typically constructed from a small piezo tube, can be designed to be very compact, simple, and rigid. The resonance frequency of these scanners is generally around tens of kHz\cite{taylor1993dynamics}, which makes it hard for the low-frequency vibrations to couple to the system. 

In contrast, generic AFM applications demand a significantly larger scan range to accommodate mesoscopic phenomena and device sizes. This range is beyond the reach of conventional piezoelectric stack actuators, which typically extend $\sim$1 $\mu$m per 100 V applied voltages per 10 mm length at low temperatures. Achieving 25 ${\mu}$m $\times$ 25 ${\mu}$m scan range requires mechanical amplification, which compromises the stiffness of the scanner. For instance, commonly used commercial scanners (Attocube) typically exhibit resonance frequencies below 200 Hz when loaded. Consequently, the AFM system becomes highly susceptible to low-frequency noise from the pulse tube, limiting the spectral noise density in the \textit{Z} axis to around 30 nm/$\sqrt{\mathrm{Hz}}$ at certain frequencies\cite{zhou2023scanning}. 

In this work, we aim to develop a cryogen-free mK-AFM with a large scan range and high \textit{Z}-resolution without requiring modifications to the dilution fridge for vibration decoupling. The microscope also incorporates a cold insertable probe to enable fast sample exchange. The primary motivation behind this tool development is to study reconfigurable oxide quantum nanoelectronics at mK temperatures. Specifically, we focus on fabricating and characterizing oxide quantum devices, such as superconducting nanowires\cite{veazey2013oxide,pai2018one} and single electron transistors\cite{
cheng2011sketched}, which can be conveniently fabricated through reconfigurable “writing” and “erasing” processes at the LaAlO$_3$/SrTiO$_3$ (LAO/STO) interface using a conductive AFM (cAFM) tip\cite{cen2008nanoscale,cen2009oxide}. These cAFM-sketched devices exhibit intriguing correlated phenomena, such as electron pairing without superconductivity\cite{cheng2015electron,cheng2016tunable}, ballistic transport\cite{cheng2018shubnikov,annadi2018quantized}, and the formation of a Pascal quantum liquid phase\cite{briggeman2020pascal}, holding great promises for quantum engineering.

We propose two design strategies to fight vibrations and heat dissipation, which are essential in this work. The first strategy is to improve the stiffness of the individual scanner stage to enable simultaneous fine scanning and coarse positioning.  Traditional mechanically amplified piezo stack actuators are used only as scanners due to their relatively low stiffness. Additional coarse \textit{XYZ}-positioners are required to position the sample precisely at a desired location. Stacking these scanners and positioners extends the mechanical loop and significantly reduces stiffness and overall performance. In contrast, we present the symmetric positioner-scanner (SyPS),\cite{CN116539920}, whose stiffness is so high that the amplified scanning mode is rigid enough to drive stick-slip motion for coarse positioning.  

The second strategy is to use fused silica glass as the major material for constructing the SyPS and microscope. Fused silica glass may be intuitively deemed too brittle for a demanding cryogenic AFM implementation. However, it has been widely used in micro-electro-mechanical systems (MEMS) due to its excellent elasticity and low internal damping\cite{hamed2023applications}. In addition, fused silica is non-magnetic, electrically insulating, and exhibits an exceptionally low thermal expansion coefficient, making it actually ideal for mK temperature applications. With a Young's modulus of $\sim$73 GPa and a density of 2.2 g/cm$^3$, the specific modulus $E/\rho$ of fused silica, where $E$ is the elastic modulus and $\rho$ is the density, surpasses titanium and beryllium copper (BeCu), materials commonly used in low-temperature stage construction. The excellent thermal insulation of fused silica can effectively isolate heat dissipation caused by piezoelectric elements. Furthermore, the low ductility of fused silica ensures greater repeatability in scanner performance by resisting plastic deformation under tensile stress. Finally, electric insulation of fused silica is preferred over titanium, which is superconducting and diamagnetic at mK temperatures. Recent advancements in glass processing techniques, such as selective laser etching (SLE), have enabled defect-free processing with features down to $\sim$200 nm under optimized conditions\cite{barbato2024femtosecond}, making glass a promising material for precision applications. In this study, we employ conventional picosecond laser-cutting technology to process fused silica. The thinnest parts of the laser-cut structure are flexure hinges, approximately 250 ${\mu}$m wide. These hinges demonstrate high reliability, with no cracking observed during a 60 ${\mu}$m $\times$ 60 ${\mu}$m scan range at room temperature and 25 ${\mu}$m $\times$ 25 ${\mu}$m at mK temperatures. More complex structures are glued together from simple laser-cut structures. 

\section{The setup}

The mK-AFM system is mounted on a top-loading probe in a cryogen-free dilution refrigerator (Leiden CF-CS81-1500M) with a base temperature of 8 mK, a cooling power of 1400 ${\mu}$W at 120 mK, and a 10 T superconducting magnet. The initial cooling is provided by a pulse tube cryocooler (Cryomech PT415) that is rigidly connected to the 50 K and 3 K plates of the refrigerator to maximize cooling efficiency. 

To characterize the vibrations of the system, we use a commercial interferometer (SmarAct Picoscale) to measure the vertical vibrations on the mixing chamber plate at room temperature. The sensor head of the interferometer is mounted on a vibration-isolated optical stand on the ground and points to the mixing chamber plate to measure the relative vibrations. 
\begin{figure}
\includegraphics[trim={0cm 0.1cm 0cm 0cm}, clip, scale=0.9]{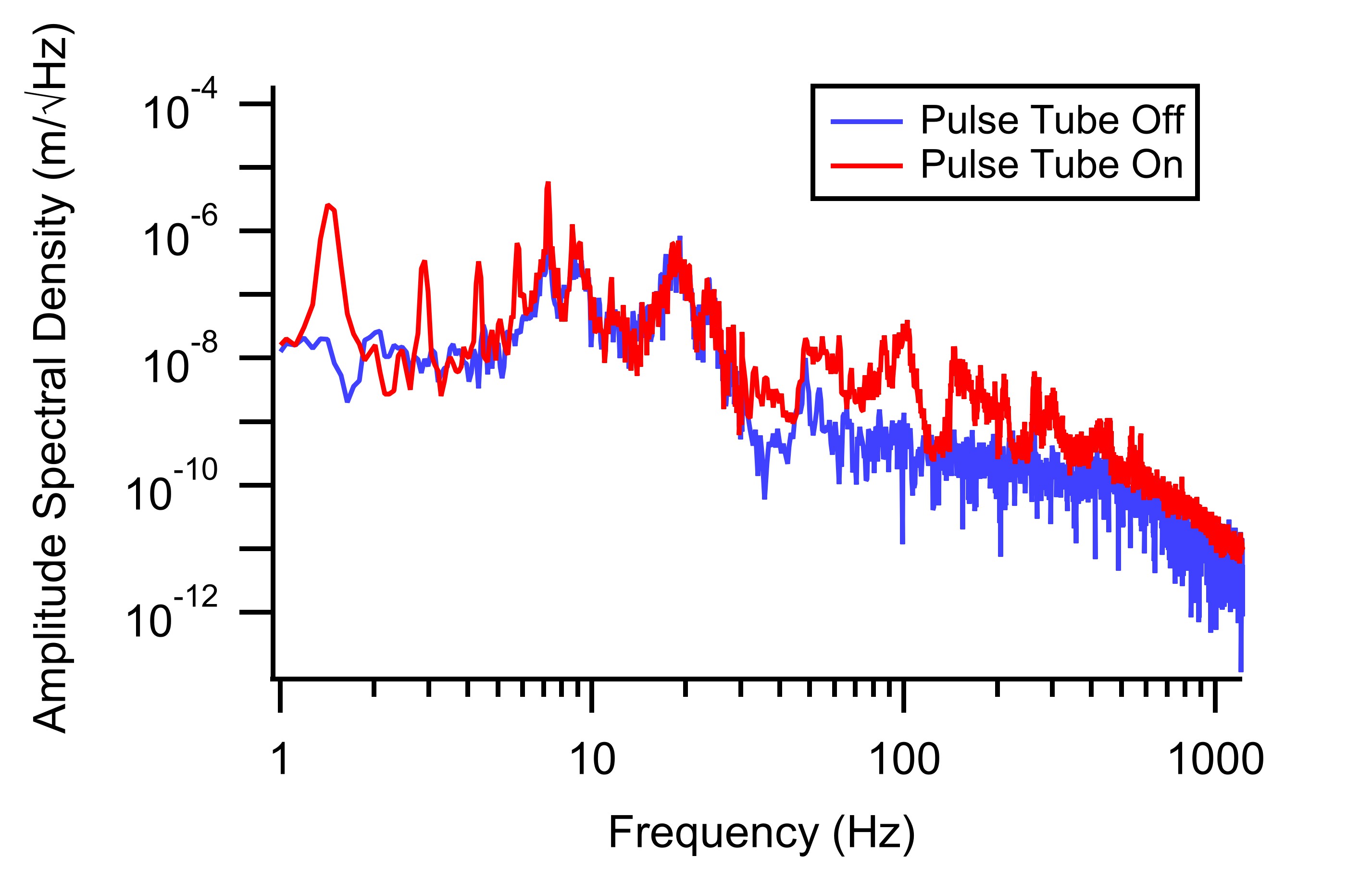}
\caption{ Vertical vibrations on the mixing chamber plate measured with the Picoscale interferometer at room temperature. The pulse frequency (1.4 Hz) and its harmonics are visible and intense in the spectrum when the pulse tube is on.}
\label{Fig.1}
\end{figure}
Figure 1 shows the vibration spectrum. When the pulse tube is on, its fundamental frequency (1.4 Hz) and harmonics are clearly visible in the spectrum, with an amplitude spectral density on the order of 10$^{-6} $ m/$\sqrt{\rm{Hz}}$ below 10 Hz. The integrated root mean square (RMS) vibration $I$ between two frequencies $f_1$ and $f_2$ is defined as
\begin{equation}
I(f_1,f_2)=\sqrt{\int{_{f_1}^{f_2}[S(f)]^2df}},
\end{equation}
where $S(f)$ is the amplitude spectral density. As a result, the measured RMS vibration in the vertical direction on the mixing chamber is 2.5 ${\mu}$m with the pulse tube on, compared to 0.8 ${\mu}$m with the pulse tube off, over the frequency range from 1 Hz to 1000 Hz.

\begin{figure*}
\includegraphics[trim={0cm 0.2cm 0cm 0.5cm}, clip, scale=0.95]{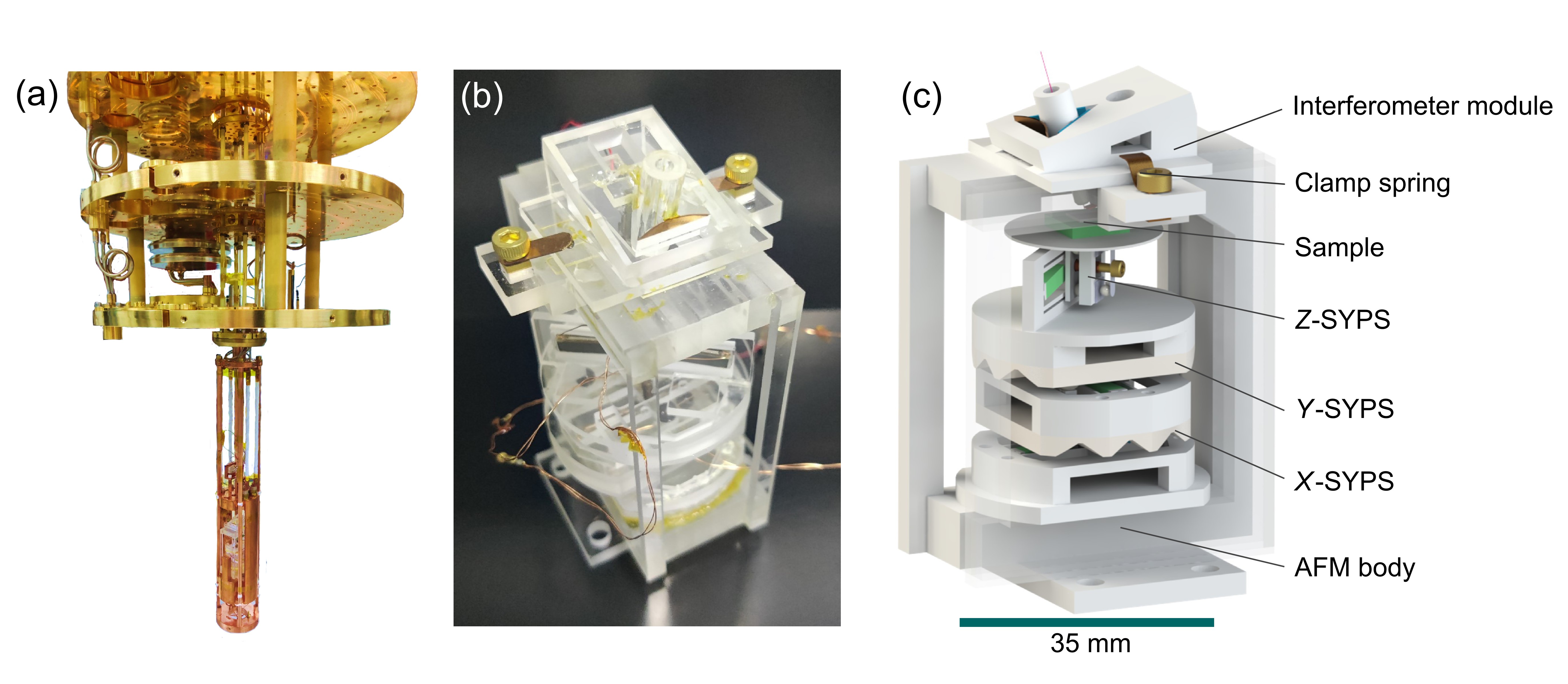}
\caption{ (a) A photograph of our mK-AFM in the Leiden dilution refrigerator. The mK-AFM can be loaded to refrigerator by a top-loading probe. (b) A photograph and (c) a 3D drawing of the glass mK-AFM.}
\label{Fig.2}
\end{figure*}
Figure 2(a) shows the glass microscope mounted in the dilution refrigerator, which consists of the triaxial scanner, infrared laser interferometer module, and the AFM body. The scanner is made by gluing 3 SyPS together using GE varnish and the interferometer module is spring-clamped to the glass body (Figs. 2(b), (c)). The microscope is softly mounted in an oxygen-free copper (OFC) shield, weighing $\sim$1.2 kg, to increase mass needed for vibration damping. The OFC shield also provides thermal anchoring and partially shields thermal radiation from the superconducting magnet. The microscope and the OFC shield are suspended from the bottom stage of the top-loading probe via four BeCu springs, ensuring the sample to remain centered within the magnet. These springs, initially $\sim$10 cm, stretched to $\sim$18 cm, are wrapped with Teflon tape which serves as a damper at cryogenic temperatures (Fig. 2(a)). The vertical resonance frequency of the OFC shield-BeCu spring system is calculated to be $\sim$1.8 Hz by $f_{\mathrm{s}}=(2\pi)^{-1}\sqrt{g/{\Delta l}}$, where $\Delta l$ is the equilibrium extension of the springs.

\subsection{\label{sec:level2}Symmetric positioner-scanner}

\begin{figure*}
\centering
\includegraphics[trim={0cm 0.3cm 0cm 0.5cm}, clip, scale=0.88]{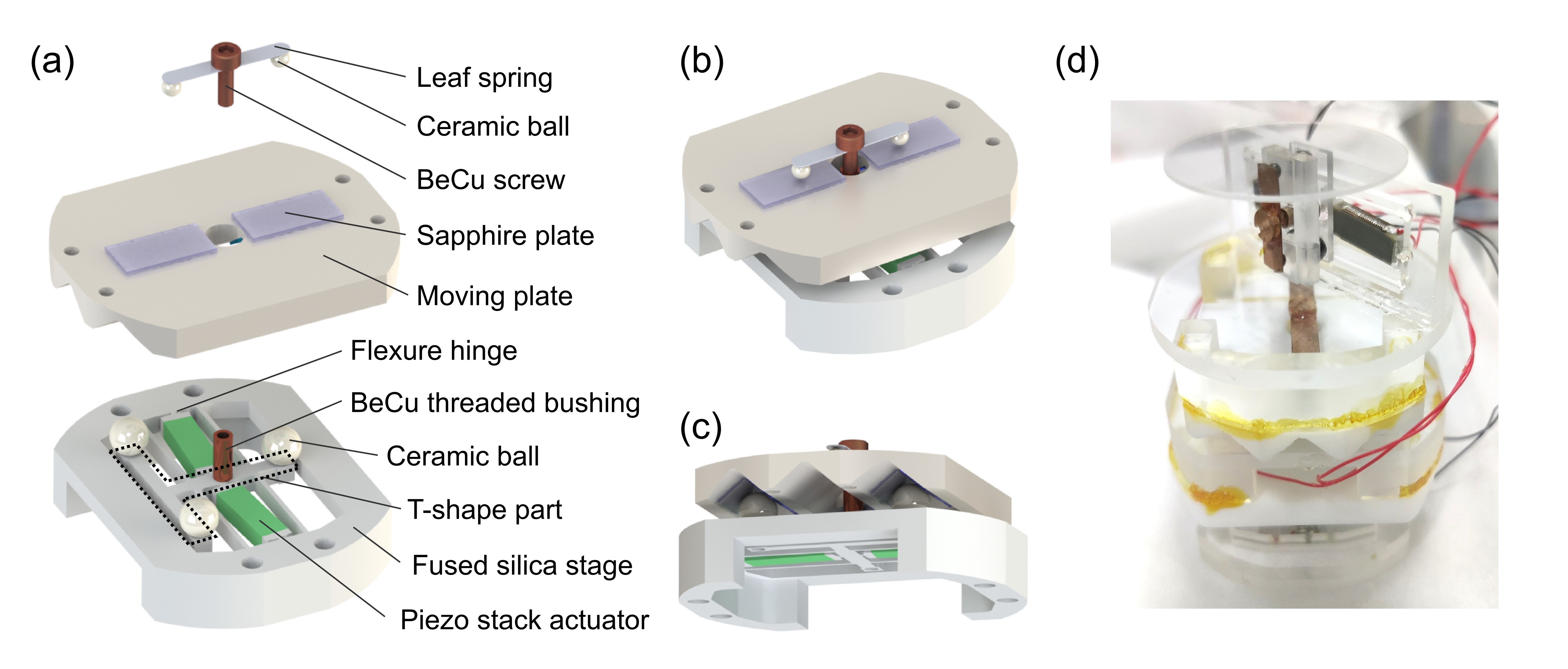}
\caption{ (a) Exploded view of the \textit{X} (or \textit{Y})-SyPS. (b,c) Schematic of assembled \textit{X} (or \textit{Y})-SyPS from two different views. (d) Photograph of the triaxial SyPS stage made of glass.}
\label{Fig.3}
\end{figure*}
Figures 3 shows the design of the \textit{X} (or \textit{Y})-SyPS, which is driven by amplified motion of two 9 mm $\times$ 3 mm $\times$ 2 mm piezo stack actuators (Physik Instrumente) individually mounted within two laser-cut flexure hinges made from fused silica. The outer ring of the fused silica stage is fixed, while the inner T-shape part is activated by amplified displacement of the piezo actuators\cite{CN116539920}. A ZrO$_2$ moving plate is mounted by contacting its wedge groove guide to 3 ceramic balls on the T-shape part for coarse position. This plate is pressed by a leaf spring fixed by a BeCu screw on the T-shape part. 

Trade-off is carefully balanced between the amplification ratio and the resonance frequency\cite{ru2016nanopositioning}, and key parameters are simulated by the finite element method (FEM). As shown in Figs. 4(a) and 4(b), fused silica has a key advantage over titanium and BeCu in achieving higher stiffness due to its higher specific modulus. At the same output range, FEM simulation indicates the fundamental resonance frequency of fused silica is with $\sim$6$\%$ and $\sim$27$\%$ higher than titanium and BeCu, respectively. 
\begin{figure*}
\centering
\includegraphics[trim={0.0cm 0.1cm 0cm 0cm}, clip, scale=0.8]{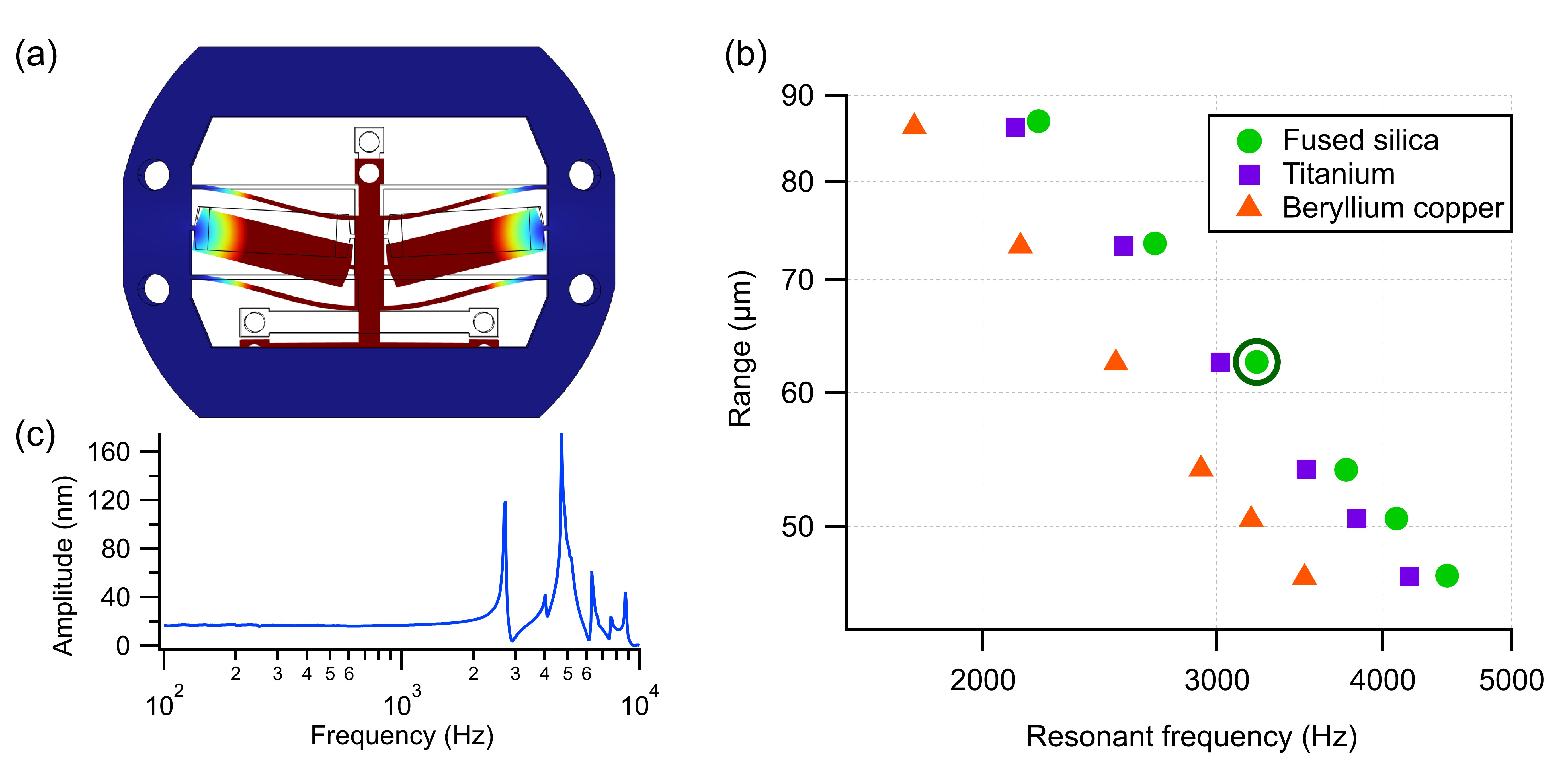}
\caption{ (a) FEM simulation of the 1$^{st}$ mechanical resonance mode of the fused silica stage using two 9 mm-long piezo stack actuators. (b) FEM simulation of the scan range at room temperature versus the 1$^{st}$ mechanical resonance frequency. The green circled point represents the selected parameters for SyPS in this work, with simulated scan range of 62 ${\mu}$m at room temperature and expected resonance frequency of 3.2 kHz. (c) Measured amplitude of the T-shape (with a 0.5 g loaded mass) versus the driving voltage frequency. The experimentally measured 1$^{st}$ resonance frequency is 2.7 kHz.}
\label{Fig.4}
\end{figure*}

The stroke of piezo stack actuators in this work is 8 $\mu$m (0 V to 120 V) at room temperature and $\sim$4 $\mu$m at cryogenic temperatures after applying larger voltages (-215 V to 215 V). By FEM, we choose a suitable scan range of 25 $\mu$m $\times$ 25 $\mu$m at cryogenic temperatures, while maintaining a high resonance frequency of $\sim$3.2 kHz. The high resonance frequency enhances the stiffness of SyPS and ensures the maximum acceleration of the T-shape part $a=(2\pi f_{\mathrm{r}})^2x_{\mathrm{T}}$, where $f_{\mathrm{r}}$ is the resonance frequency and $x_{\mathrm{T}}$ is the displacement of the T-shape, to drive the coarse positioning mode by stick-slip motion.

Figure 4(c) shows the measured  resonance frequency of the fused silica SyPS, obtained by sweeping the driving voltage frequency with a fixed 2 V amplitude and detect the displacement by the Picoscale interferometer. With a 0.5 g loaded mass (corresponds to the 3 ceramic balls, a threaded bushing, and a mirror on the T-shape part), the first resonance peak occurs at 2.7 kHz, which is still high enough to enable coarse positioning.

The \textit{Z}-SyPS has the same design as the \textit{X} (or \textit{Y})-SyPS, except for a moving plate of reduced height and weight. The \textit{Z}-SyPS can support a load greater than 12 g. We glue three SyPSs together with GE varnish to form a triaxial SyPS stage as shown in Fig. 3(d). 

\subsection{Infrared laser interferometer module}

We adopt a fiber-cantilever interferometer design as in Ref. \onlinecite{rugar1989improved} by using balanced detection and pulsed modulation for cantilever displacement sensing (Fig. 5(a)). Such a design is simple, highly sensitive and has been applied in many home-made AFM fabrications\cite{kracke1996ultrahigh,von2016understanding,karci2014design}. The main challenge of the interferometer at cryogenic temperatures is laser alignment and heat dissipation. In our all fiber setup, the 1550 nm laser is first split, with half of the light routed to fiber-cantilever cavity for interference signal, and the other half serves as reference light to cancel laser noise through balance detection (Thorlabs PDB450C). The initial 1 mW laser power is pulse-width modulated (PWM) with a 3.3$\%$ duty cycle (100 $\mu$s on-time per cycle) to minimize heating. The mixing chamber plate temperature remains unaffected compared to when the laser diode is off. 

\begin{figure*}[t]
\centering
\includegraphics[trim={0.0cm 0.8cm 0.0cm 1cm}, clip, scale=0.9]{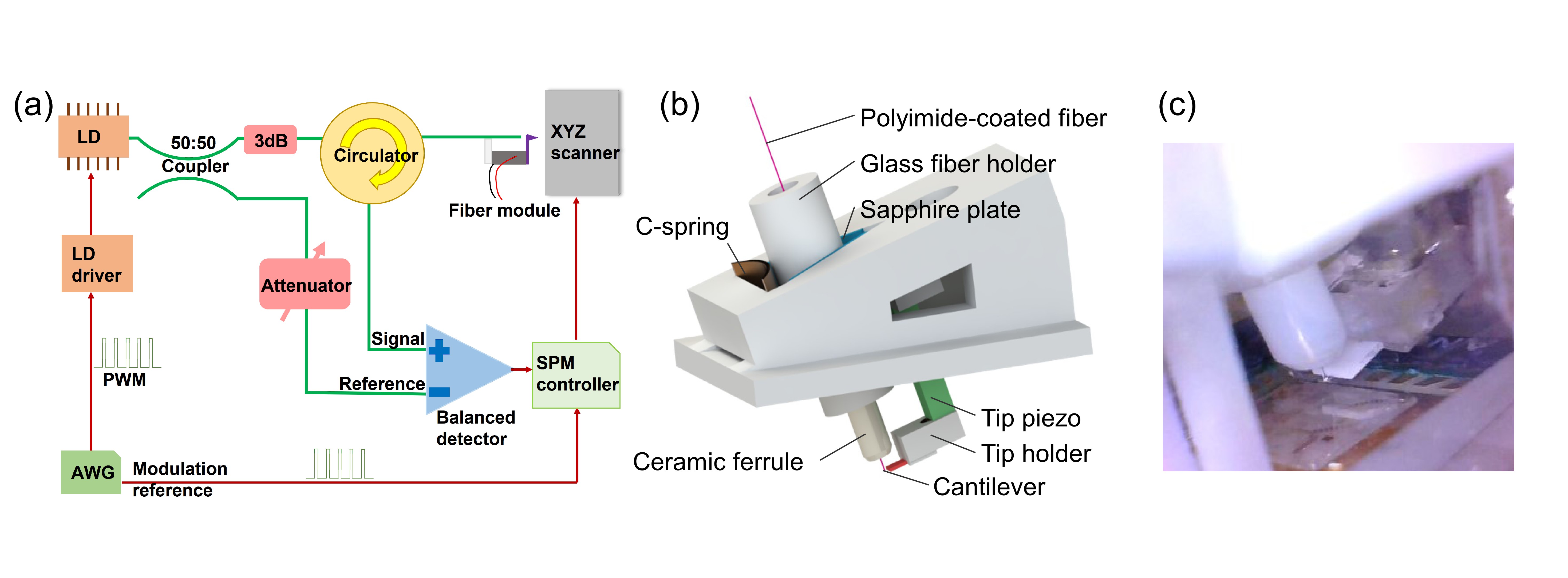} 
\caption{ (a) Schematic of the laser interferometer setup. (b) Drawing and (c) Photograph of the interferometer module.}
\label{Fig.5}
\end{figure*}
Figure 5(b) shows the glass interferometer module which maintains alignment from room temperature to 100 mK by taking advantage of the extremely low thermal expansion coefficient of fused silica. Additionally, the perfect electrical insulation of fused silica glass makes it suitable for common AFM modes like cAFM, piezoresponse force microscopy, scanning gate microscopy, and electrostatic force microscopy. Its lightweight design also contributes to excellent mechanical stability, while its high transparency facilitates sample observation and tip coarse positioning at room temperature.

The AFM cantilever is spring-clamped onto a glass tip holder, which is angled at 15$^\circ$ and installed with a tip piezo to adjust fiber-cantilever spacing. The polyimide-coated fiber is held by a glass holder and sandwiched between a polished BeCu C-spring and two sapphire plates glued to a triangular hole. The fiber end is coarse-approached to within 100 $\mu$m of the cantilever by manually adjusting the tip holder and fiber holder. The fiber-cantilever alignment is verified by observing the shadow of cantilever on a infrared laser viewing card and inspecting the spacing under an optical microscope. Once aligned, the interferogram is measured by linearly extending the tip piezo. In tapping mode AFM, the tip piezo also excites the cantilever with a nanometer-scale amplitude by applying a small AC voltage.

\subsection{Body structure, cabling, and thermal anchoring}

The body of our mK-AFM is also made of fused silica glass to ensure maximum stiffness. We assemble laser-cut glass plates and glue together to create a low-cost 3D body frame. The simulated resonance frequency of the AFM body exceeds 1 kHz.

We use soft stainless-steel coaxial cables (New England Wire Technologies N12-50F-257-0) for all the high-voltage lines for piezo actuators and low-voltage lines for transport measurements. These cables and the fiber are routed from room temperature to the mixing chamber plate in the top-loading probe, with careful thermal anchoring at each intermediate temperature plate using OFC clamps and silver paint.  At the mixing chamber plate on the probe, we switch to soft silver-coated copper coaxial cables to ensure good thermal conductivity. The contact between the probe and dilution refrigerator plates is established by a pneumatically activated clamping system, a characteristic design of Leiden refrigerators.

The power dissipation generated by the SyPS stage is given by $P=2\pi f_{\mathrm{V}}CV^2\tan {\delta}$, where $C$ and tan$\delta$ are the capacitance and the dissipation factor of the piezo actuators. With amplified stick-slip design feature of SyPS, the voltage \textit{V} required to drive is small ($\sim$30 V) and driving frequency $f_{\mathrm{V}}$ is also reduced ($\sim$10 Hz) at $T=$100 mK , which significantly reduces power dissipation. Together with the excellent thermal insulation of fused silica, we find that neither coarse positioning nor fine scanning affects the temperature ($<100$ mK) which is measured by the RuO$_2$ thermometer mounted on the OFC shield.

\section{Performance}

The typical cooling rate we use is $\sim$10 K/h. The glass AFM has been cycled five times without any cracks and has shown stable performance throughout our experiments.

\begin{figure}[b]
\includegraphics[trim={0.0cm 1.3cm 0.0cm 0.2cm}, clip, scale=0.8]{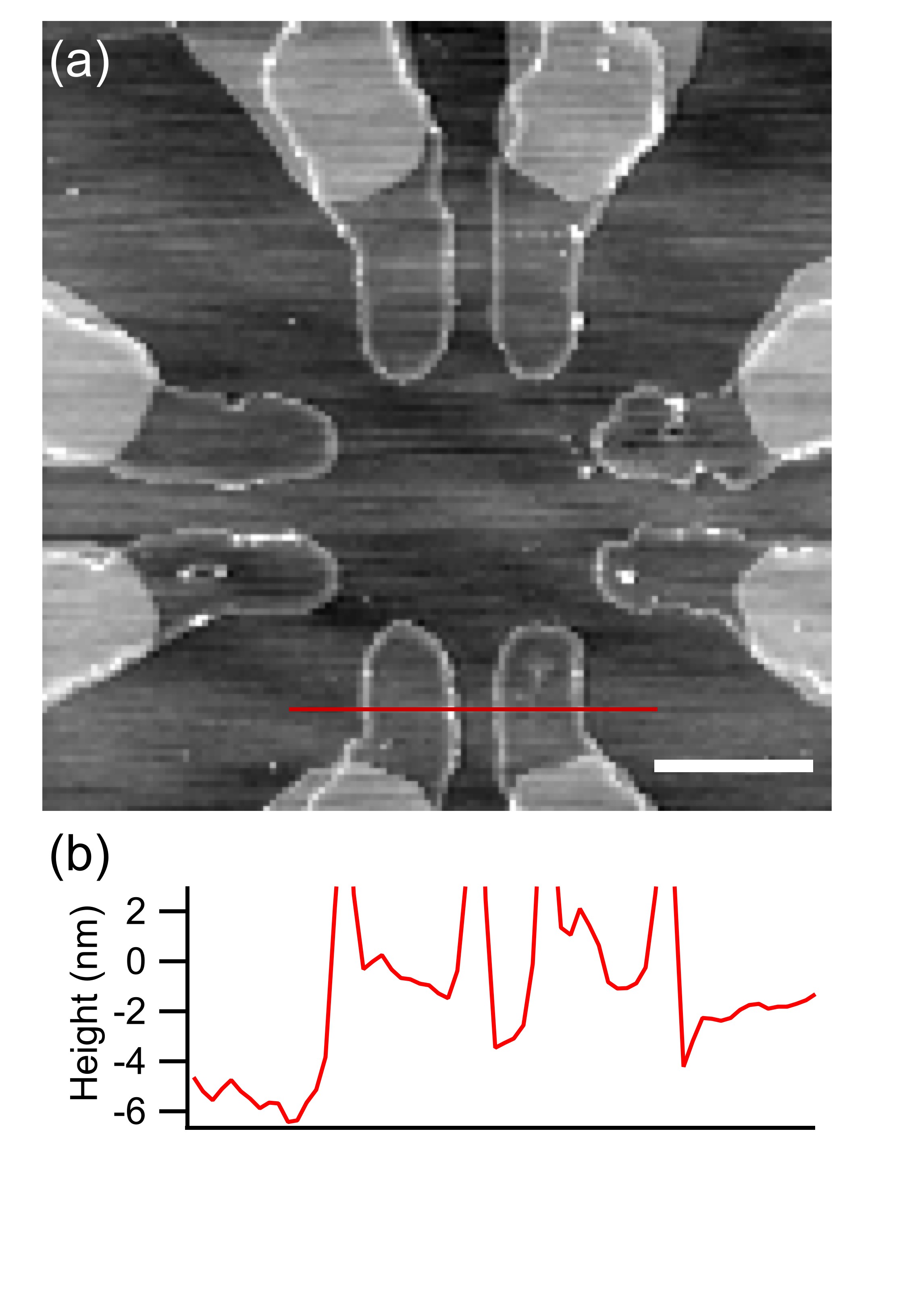}
\caption{(a) Topography scan of an LAO/STO canvas at $T=100$  mK with pulse tube on. (b) The line profile along the red cross line. Scale bar, 5 $\mu$m.}
\label{Fig.6}
\end{figure}
The essential scanning functions of our mK-AFM is controlled by an RHK R9plus controller. Figure 6 shows the contact-mode topographic imaging of the interfacial electrodes of an LAO/STO sample, acquired at $T=100$ mK with a scan rate of 0.25 Hz, a bandwidth of 100 Hz, and a pixel resolution of 128 $\times$ 128. These gold electrodes, which define a cAFM writing "canvas", are fabricated using Ar-ion milling and \textit{in-situ} sputter deposition techniques to contact the interface\cite{cheng2011sketched}. They stick $\sim$5 nm above the surface and are higher on the edge due to fabrication imperfections. The whole image is clean with pulse tube on, suggesting remarkable resilience on environmental vibrations.  

To characterize the \textit{Z} noise in contact mode AFM imaging under normal operating conditions of the dilution refrigerator, we performed a 256 $\times$ 256-pixel scan at 0.5 Hz rate and 100 Hz bandwidth over a 2 nm $\times$ 2 nm area on a flat sample surface, where the spatial profile can be neglected. As shown in Fig. 7, the standard deviation ($\sigma$) of a Gaussian fit is $\sim$0.6 nm ($\sim$4 ms pixel time).
\begin{figure}[b]
\centering
\includegraphics[trim={0.0cm 0.0cm 0.0cm 0.5cm}, clip, scale=0.80]{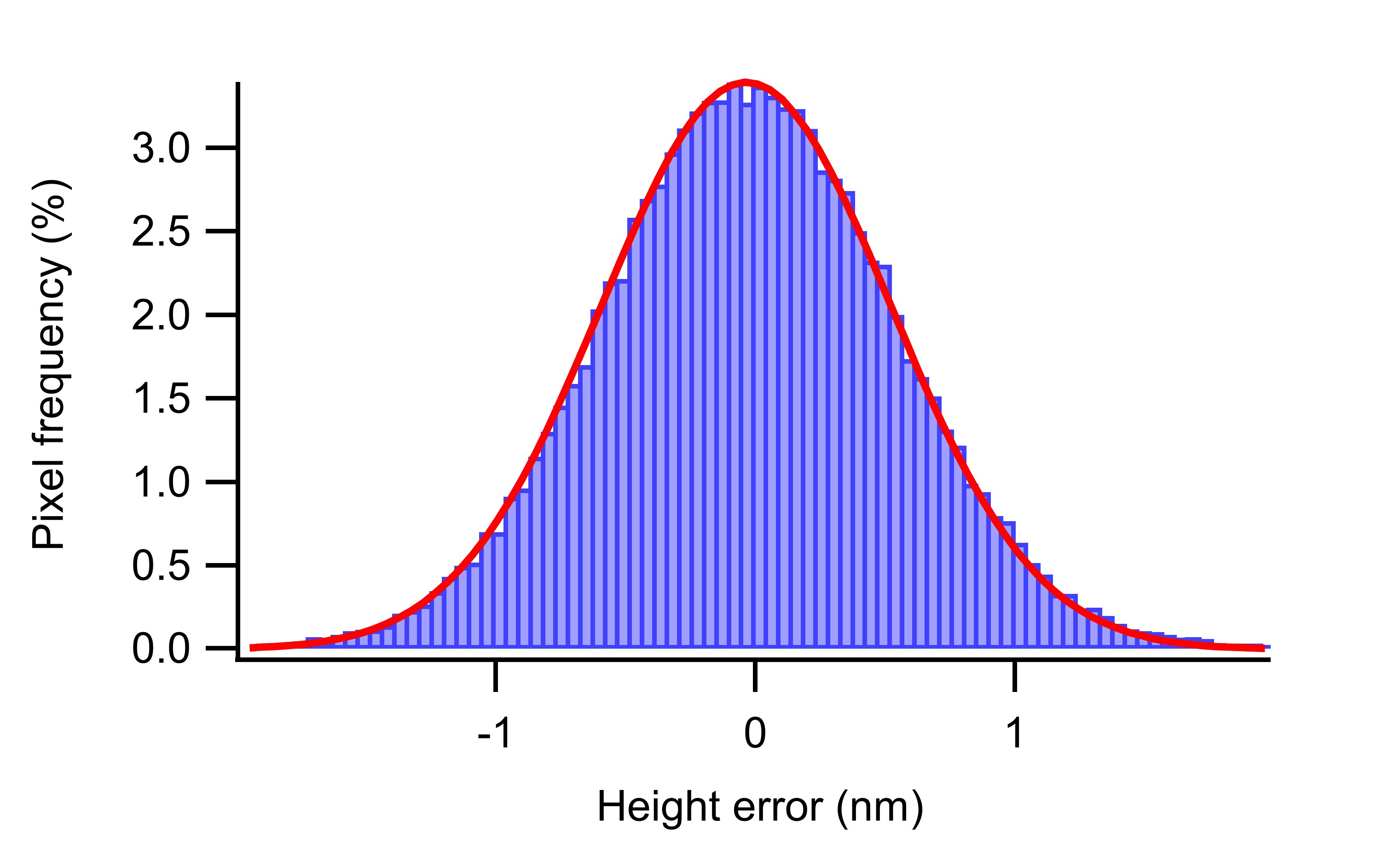}
\caption{Height error measured from a 256 $\times$ 256-pixel noise scan over a 2 nm $\times$ 2 nm area.}
\label{Fig.7}
\end{figure}
 The lateral resolution of the contact mode AFM scanning depends on the sharpness of the tip, which may become blunt during a long experiment at low temperatures.

\section{Conclusion}

In conclusion, we have presented a low cost AFM design made of glass, featuring superior performance in a cryogen-free dilution refrigerator owing to the added stiffness and reduced thermal dissipation. The benefits mainly come from the stiff symmetric positioner-scanners design and the excellent specific modulus of glass. The sub-nm stability, mK operating temperatures and the simultaneous transport measurement capability grant us with the needed tool to explore reconfigurable oxide quantum nanoelectronics. Further improvements can be made by increasing the stiffness of \textit{Z}-SyPS by choosing a smaller amplification. Additionally we aim to design a compact, quasi-zero-stiffness vibration isolator to replace the BeCu springs to lower vibration levels. Finally, 3D processing of glass structures by using SLE technology may enhance the overall performance of the glass mK-AFM. 

\begin{acknowledgments}
This work was supported by the CAS Project for Young Scientists in Basic Research (YSBR-100), the National Key Research and Development Program of China (2024YFA1409500), and the Fundamental Research Funds for Central Universities (KY2030000160 and WK3540000003).
\end{acknowledgments}

\section*{Data Availability Statement}

The data that support the findings of this study are available from the corresponding author upon reasonable request.

\nocite{*}
\bibliographystyle{unsrt}
\bibliography{reference}

\end{document}